\documentclass[letterpaper, 10 pt, conference]{ieeeconf}  

\IEEEoverridecommandlockouts                              

\usepackage{graphics} 
\usepackage{graphicx}
\usepackage{hyperref}
\usepackage{epsfig} 
\usepackage{amsmath} 
\usepackage{amssymb}  
\usepackage{cite}
\usepackage{booktabs}   
\usepackage{makecell}   
\usepackage{xcolor}

\title{\LARGE \bf
	Towards Automated EEG-Based Epilepsy Detection Using Deep Convolutional Autoencoders
}
\author{Annika Stiehl$^{1}$*, Nicolas Weeger$^{1}$, Christian Uhl$^{1}$, Dominic Bechtold$^{1}$, Nicole Ille$^{2}$ and Stefan Geißelsöder$^{1}$
	\thanks{*This work was supported by Bavarian Ministry of Economic Affairs, Regional Development and Energy (StMWi, Funding number: DIK-2307-0007// DIK0536/01) and by the German Federal Ministry of Education and Research (BMBF, Funding number:
		05M20WBA).}
	\thanks{$^{1}$University of Applied Sciences Ansbach, Residenzstr. 8, 91522 Ansbach, Germany,
		{\tt\small annika.stiehl@hs-ansbach.de}}%
	\thanks{$^{2}$BESA GmbH, Freihamer Str. 18, 82166 Gräfelfing, Germany
		{\tt\small nille@besa.de}}%
}

\begin{document}
	\maketitle
	\thispagestyle{empty}
	\pagestyle{empty}
		
	\begin{abstract}		
		Epilepsy is one of the most common neurological disorders. This disease requires reliable and efficient seizure detection methods. Electroencephalography (EEG) is the gold standard for seizure monitoring, but its manual analysis is a time-consuming task that requires expert knowledge. In addition, there are no well-defined features that allow fully automated analysis. Existing deep learning-based approaches struggle to achieve high sensitivity while maintaining a low false alarm rate per hour (FAR/h) and lack consistency in the optimal EEG input representation, whether in the time or frequency domain.
		To address these issues, we propose a Deep Convolutional Autoencoder (DCAE) to extract low-dimensional latent representations that preserve essential EEG signal features. The ability of the model to preserve relevant information was evaluated by comparing reconstruction errors based on both time series and frequency-domain representations. Several autoencoders with different loss functions based on time and frequency were trained and evaluated to determine their effectiveness in reconstructing EEG features. Our results show that the DCAE model taking both time series and frequency losses into account achieved the best reconstruction performance. This indicates that Deep Neural Networks with a single representation might not preserve the relevant signal properties. This work provides insight into how deep learning models process EEG data and examines whether frequency information is captured when time series signals are used as input.
	
	\end{abstract}

	\section{INTRODUCTION}
		
		According to the World Health Organization (WHO), epilepsy is a worldwide chronic brain disorder affecting around 50 million people of all ages \cite{WHO2024}. It is characterized by recurrent, unprovoked seizures, which in aroud 30 \% of cases can not be controlled with anti-epileptic drugs (AEDs)~\cite{HerbozoContreras2024}. Seizures can occur at any time, with or without aura, and can have a significant impact on the patient's life~\cite{Ma2024}. The diagnosis of epilepsy is typically made on the basis of the amplitude and waveform of Electroencephalography~(EEG), a technique that is widely regarded as an effective tool for monitoring and recording brain activity~\cite{Ebner2011, Mansilla2024}. Using EEG, three distinct stages of brain activity can be identified: the ictal, interictal, and preictal phases, which provide information for seizure detection and monitoring.
		
		The need for reliable and automated seizure detection is driven by the limitations of manual EEG analysis, which is a time-consuming task and requires specialized expertise \cite{Dan2024, Sousa2024, Zhou2024}. Nevertheless, accurate seizure detection is critical for effective patient management. Automated systems aim to reduce the workload of clinicians and enable better monitoring of patients \cite{Dan2024, DeBrabandere2023, Kerr2024}. 
		
		\begin{figure*}[t!]
			\centering
			\includegraphics[width=1\textwidth]{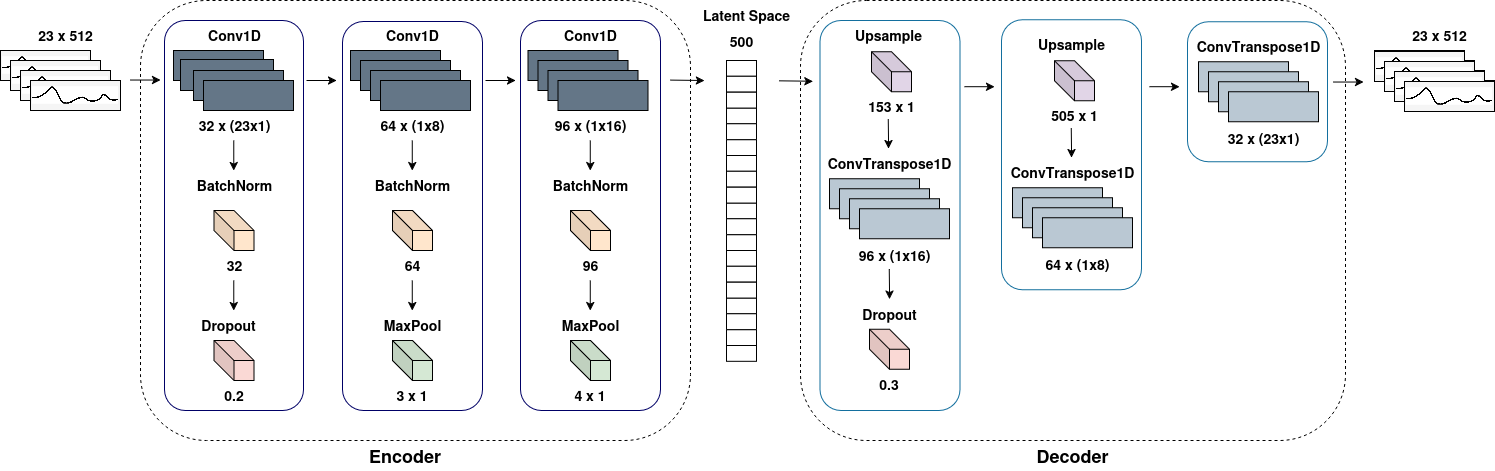}
			\caption{Network architecture of the Deep Convolutional Autoencoder with 23 EEG channels as input.}
			\label{fig:AE}
		\end{figure*}
		
		EEG signals are inherently non-linear, non-stationary, and dynamic, presenting challenges for traditional algorithms to effectively capture their complex characteristics \cite{GowthamReddy2024, Li2024, Zhou2024}. Moreover, the presence of artifacts caused by muscle movements, blinking, or external factors can obscure or mimic seizure patterns, complicating automatic detection~\cite{Pedoeem2022}. These artifacts highlight the importance of robust preprocessing and artifact-handling strategies in seizure detection pipelines.
		
		Another critical challenge lies in the scarcity and imbalance of EEG datasets. Large, annotated datasets of recorded seizures are often unavailable due to privacy concerns. In addition, the available datasets often have a significant imbalance, containing much more non-seizure EEG data than seizure data \cite{Dan2024, Hassan2024, Amin2016}. This imbalance can lead to biased models that are less sensitive to seizures and more prone to misclassification.
		
		In addition to data challenges, the subjective nature of seizure definitions introduces variability in labeling. Even experienced clinicians often disagree on the exact onset and termination of a seizure or whether a particular pattern constitutes a seizure at all \cite{Pedoeem2022}. This low inter-rater accuracy complicates the creation of consistent and reliable training datasets, ultimately affecting model performance.

		Finally, the lack of generalizability in many existing seizure detection models remains a significant limitation. These models are often highly patient-specific and fail to generalize to new individuals \cite{Peh2023, Tsai2023}. Developing patient-independent models that can accommodate the variability in seizure characteristics across different populations is a key research challenge.
		
		Automated seizure detection has been studied since the 1970s, with Gotman pioneering early computational approaches for EEG analysis \cite{Gotman1976}. With the increasing use of machine learning in EEG analysis, various techniques, including feature engineering and automated classification, have been explored to improve the identification of epileptic patterns and seizure-related abnormalities \cite{Stiehl2023, DeBrabandere2023, Tong2024}. In recent years, Convolutional Neural Networks (CNNs) have shown strong performance in EEG analysis and seizure detection, effectively extracting features from EEG signals \cite{Gao2020, SilvaLourenco2021}. However, determining the optimal EEG representation remains a challenge. Autoencoders (AEs) have emerged as an alternative demonstrating their potential for unsupervised feature learning and false alarm reduction \cite{Wen2018, Takahashi2020}. More recently, Fu et al. \cite{Fu2024a} proposed a self-supervised masked autoencoder for EEG, improving representation learning. Despite these advancements, achieving high sensitivity while maintaining a low false alarm rate remains an open challenge.
		
		The contribution of this paper is the development of a novel Deep Convolutional Autoencoder (DCAE) architecture (Fig. \ref{fig:AE}) that incorporates a frequency-domain component into the reconstruction loss. By explicitly considering frequency information, the model focuses on capturing relevant spectral features in the latent space while reconstructing the time series. This approach aims to improve the representation of seizure-related features and improve the generalizability of the model for downstream tasks such as seizure classification.
		
	\section{METHODS}
		
		For all steps Python is used as the programming language, leveraging several key libraries \footnote{Code is available at: \url{https://github.com/HS-Ansbach-CCS/embc_25_dcae}}. The edfio library is utilized for handling EDF files, while numpy, pandas, and scipy are employed for data type manipulation and processing. For data visualization we use matplotlib, and pytorch to implement machine learning tasks. 
		
		\subsection{Dataset}
			
			We use an open-access available scalp EEG dataset from the Temple University Hospital (TUH) \cite{Shah2018}. The TUH EEG Seizure Corpus (TUSZ) is divided into three different datasets: train, dev, and eval. The train and dev datasets are used as intended for training the machine learning algorithm and optimizing its hyperparameters. The eval dataset will be used to evaluate the finalized model on unseen data \cite{Golmohammadi2021}. In total, the dataset contains 7361 EEG recordings from 675 different patients. An overview of the dataset, including its demographics and characteristics, is provided in Table \ref{table_tusz} and Figure \ref{fig:smprate}. 
			
			The demographic distribution in Tab. \ref{table_tusz} shows an equal distribution of male and female patients, but only half of the patients have a seizure event in their recordings. The number of annotated seizure events is high compared to other open-access scalp EEG datasets, but still seizures are clearly underrepresented. All recordings are stored in the European Data Format (EDF).
			
			\begin{table}[t!]
				\caption{Overview of the TUSZ dataset}
				\label{table_tusz}
				\centering
				\small 
				\setlength{\tabcolsep}{4pt} 
				\renewcommand{\arraystretch}{0.9} 
				\begin{tabular}{l c c c c}
					\makecell[l]{Dataset \\ Properties} & Train & Dev & Eval & Total\\
					\midrule
					\midrule
					Files & 4664 & 1832 & 865 & 7361\\
					\midrule
					\makecell[l]{Patients \\ (with seizure)} & 579 (208) & 53 (45) & 43 (34) & 675 (287)\\
					\midrule
					Female & 301 & 31 & 23 & 355\\
					\midrule
					Male & 278 & 22 & 20 & 320\\
					\midrule
					\makecell[l]{Total duration \\ (hours)} & 910.34 & 435.55 & 127.70 & 1473.59\\
					\midrule
					\makecell[l]{Total seizure \\ duration (hours)} & 48.65 & 19.96 & 7.57 & 76.18\\
				\end{tabular}
			\end{table}
			
		\subsection{Autoencoder}
			
			An autoencoder (AE) is a neural network that consists of an encoder and a decoder \cite{Goodfellow2016}. It represents an unsupervised learning method that learns to encode and reconstruct input data without requiring explicit labels for training. The encoder part maps the input $x\in\mathbb{R}^{C\times T}$ to a lower-dimensional latent space vector $z\in \mathbb{R}^d$. Here, $C$ represents the number of channels, $T$ denotes the number of time points, and $d$ corresponds to the dimensionality of the latent space. The decoder reconstructs the input $\hat{x}\in\mathbb{R}^{C\times T}$ from the latent vector $z$.
			
			Mostly the dimension of the latent space is less than the input dimensions. So, AEs are able to compress data to meaningful features. Since autoencoder learn features tailored to the training data, their application is limited to the data distribution used during training, distinguishing them from standard compression methods.
			
			The AE is trained to minimize the reconstruction loss between the input and the reconstructed signal from the decoder, which can be the Mean Absolute Error (MAE):
			
			\begin{equation}
				\mathcal{L} = \frac{1}{n} \sum_{i = 1}^{n} |x_i - \hat{x}_i|.
			\end{equation}
			
			Since the MAE is mostly not zero, the compression or dimension reduction is incurred with loss. AEs can be used as dimension reduction method to extract meaningful features for later analysis of the data. 
			
			The architecture of the Deep Convolutional Autoencoder (DCAE) is illustrated in Fig. \ref{fig:AE}. The input EEG signal has an initial dimension of 23 channels × 512 time points. The encoder consists of three convolutional blocks, each containing a convolutional layer with different kernel sizes, followed by batch normalization. In the first block a dropout layer with a dropout rate of 20\% is added. The second and third convolutional blocks contain a maximum pooling layer. The extracted features are then mapped to a fully connected latent space with 500 neurons. The decoder mirrors the encoder structure, utilizing three deconvolutional blocks with upsampling and dropout layers to reconstruct the original EEG signal.
			
		\subsection{Data Preprocessing and Preparation}
			
			Before applying any machine learning algorithms, the EEG data undergoes preprocessing and preparation. This process consists of multiple steps: data cleaning, resampling, filtering, remontaging, windowing, scaling, and data augmentation. Each of these steps is detailed in the following sections.
			
			\paragraph{Data Cleaning}
			As part of data cleaning, we remove the intracranial electrodes SP1 and SP2. Additionally, EEG channels with unknown locations or unclear labeling are discarded. Auxiliary physiological signals, such as respiration or electrocardiogram (ECG) data, are also excluded from the recordings.
			
\begin{figure}[t!]
				\centering
				\includegraphics[width=0.7\linewidth]{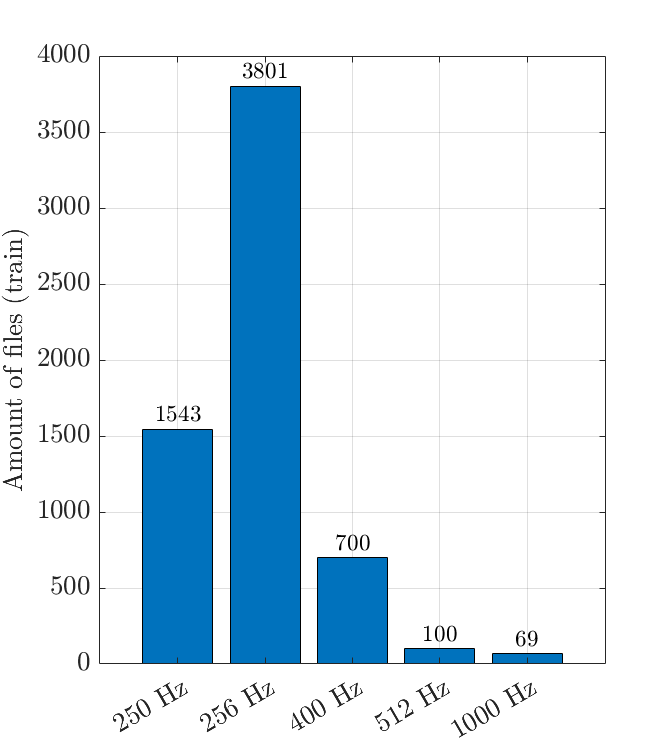}
				\caption{Initial sampling rates of the EEG recordings from the train dataset.}
				\label{fig:smprate}
			\end{figure}

			\paragraph{Resampling}
			As most TUSZ EEG signals are sampled with 256~Hz (see Fig. \ref{fig:smprate}), we choose 256~Hz as our base sampling rate. As shown in Fig. \ref{fig:smprate} some files required upsampling (from 250 Hz) or downsampling. For this step, cubic spline interpolation was applied.
			
			\paragraph{Filtering}
			A bandpass filter (0.5~Hz~-~70~Hz) is applied to remove unwanted artifacts. Since the data is recorded in the USA, we apply a 60~Hz notch filter to suppress power line interference. The high-pass filter is designed as a first-order Butterworth filter with a 6~dB/oct slope, while the low-pass filter is a fourth-order (24~dB/oct) Butterworth filter. The notch filter was implemented as a second-order Infinite Impulse Response (IIR) filter with a 2~Hz bandwidth.
			
			\paragraph{Montage Standardization}
			To ensure consistency across recordings, we selected the average montage and chose the EEG channels based on their frequency of occurrence in the dataset. The final 23 scalp electrodes included are:
			\textit{FP1, FP2, F3, F4, C3, C4, P3, P4, O1, O2, F7, F8, T7, T8, P7, P8, Fz, Cz, Pz, A1, A2, T1, T2.}
			
			\paragraph{Windowing and Plausibility Checks}
			EEG signals were segmented into 2-second windows. Additionally, a plausibility check was performed to filter implausible windows:
			\begin{itemize}
				\item windows with a standard deviation $>$ 5000 $\mu$V and
				\item windows where more than 8 channels had a standard deviation $<$ 0.01 $\mu$V (almost zero activity) 
			\end{itemize}
			were removed. No other additional artifact correction was applied.
			
			\paragraph{Feature Scaling}
			Similar to the RobustScaler from~\cite{RobustScaler}, a Histogram Scaler is implemented to normalize the EEG data. The scaler was trained on the training set, creating an expandable histogram to determine the median, 5th~percentile, and 95th~percentile values. The scaling function was defined as:
			
			\begin{equation}
				scaledValue = \dfrac{value - median}{percentile_{95} - percentile_{5}} .
			\end{equation}
			For each electrode channel a separate histogram was examined and scaled independently. Prior analysis showed varying mean and standard deviation values across channels due to different brain activity levels. After scaling, the data is clipped to the range [-1, 1] to prevent extreme values influencing the model and to ensure numerical stability during training. 
			
			\paragraph{Data Augmentation}
			To enhance model generalization, time jitter and electrode flipping were applied:
			\begin{itemize}
				\item Electrode flipping: Left and right brain hemisphere electrodes were swapped and
				\item Time Jitter: Windows were extracted with a 50\% overlap (i.e., each window contained 1 second of unique signal).
			\end{itemize}
			The time jitter was performed during window extraction, while electrode flipping was applied on-the-fly during model training.
			
		\subsection{Model Training} \label{model_training}
			\begin{figure}[t!]
				\centering
				\includegraphics[width=1\linewidth]{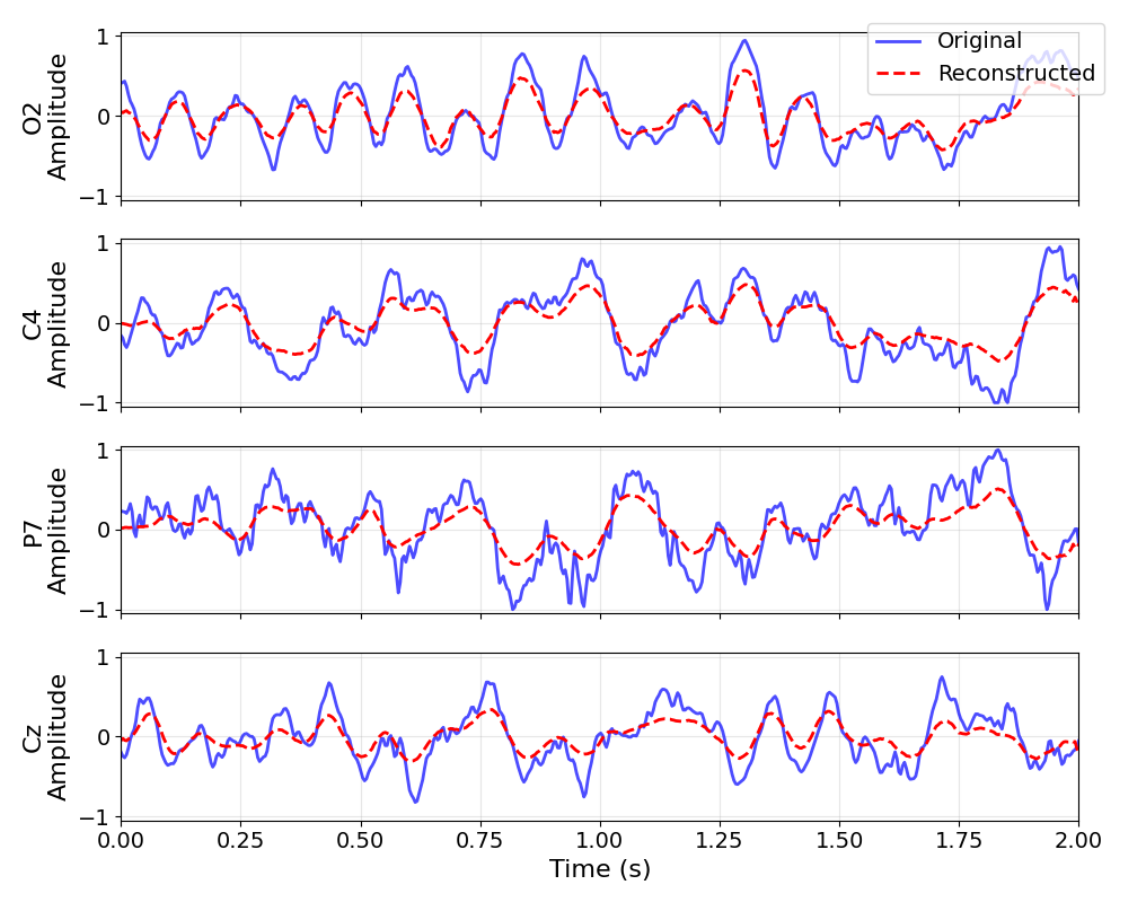}
				\caption{Example of time series reconstruction of a seizure event using DCAE\textsubscript{ts-loss} for selected channels.}
				\label{fig:eeg}
			\end{figure}
			
			The models are trained on an NVIDIA A40 GPU using a batch size of 256 samples. Optimization is performed using the Adam optimizer \cite{Kingma2014} with a learning rate of 0.001. Each model is trained 50 epochs in total.
			
			The model was optimized using the Mean Absolute Error (MAE) as the primary loss function. Different loss configurations were evaluated to balance reconstruction quality in both time and frequency domains.
			
			The first approach considers a pure time series reconstruction loss $\mathcal{L}_{TS}$, where the MAE is computed as the absolute difference between the reconstructed and original time series signals.
			
			To enhance frequency-domain representation, a combined loss incorporating the Fourier transform~(FT) and time series~(TS) reconstruction was implemented. Alpha and beta frequency bands (8-30~Hz) are known to carry important physiological and pathological information in EEG recordings, particularly in the context of seizure detection and characterization \cite{Niedermeyer2017}. Similar to the approach taken in \cite{UrbinaFredes2024}, who focused exclusively on alpha and beta waves for seizure detection, this configuration also uses only these frequency components. The corresponding loss function is defined as:
			
			\begin{equation} \label{loss_ts_ft}
				\mathcal{L} = 20 \cdot \mathcal{L}_{FT} + 1 \cdot \mathcal{L}_{TS}
			\end{equation}
			
			Both the original and reconstructed Fourier transforms are normalized to the 99th percentile of the original signal's Fourier transform. This normalization reduces the impact of extreme values and ensures a more stable loss calculation, preventing outliers from dominating the optimization process. 
			
			The multiplier 20 for the $\mathcal{L}_{FT}$ was selected so that both error components are approximately equally represented in the final loss. 
			
			Another approach incorporates the Short-Time Fourier Transform (STFT) into the loss function to capture localized frequency variations. The STFT was computed using a window size of 64 time points and a hop length of 8, resulting in an 87.5\% overlap between consecutive windows. Similar to the Fourier transform, both the original and reconstructed spectrograms were normalized to the 99th percentile of the original spectrogram to provide a stable loss computation. The final loss function is given by:
			
			\begin{equation} \label{loss_ts_spectro}
				\mathcal{L} = 20 \cdot \mathcal{L}_{STFT} + 1 \cdot \mathcal{L}_{TS}.
			\end{equation}
			
			To ensure comparability in the validation step, a frequency-only reconstruction error was also considered. In this approach, only the magnitude of the Fourier transform is used for evaluation, with all frequency components included and no additional weighting applied. Before computing the error, the Fourier transform is normalized to the 99th percentile of the original signal. Similar to the $\mathcal{L}_{FT}$ factor, the $\mathcal{L}_{STFT}$ was also set to 20. The reason for this is that both error components should have a comparable contribution to the total error $\mathcal{L}$. 
			
			These different loss configurations were analyzed to assess their impact on preserving relevant EEG features in the latent space, with a particular focus on the frequency-domain characteristics relevant for epilepsy detection.
		
	\section{RESULTS}
		\begin{figure}[t!]
			\centering
			\includegraphics[width=1\linewidth]{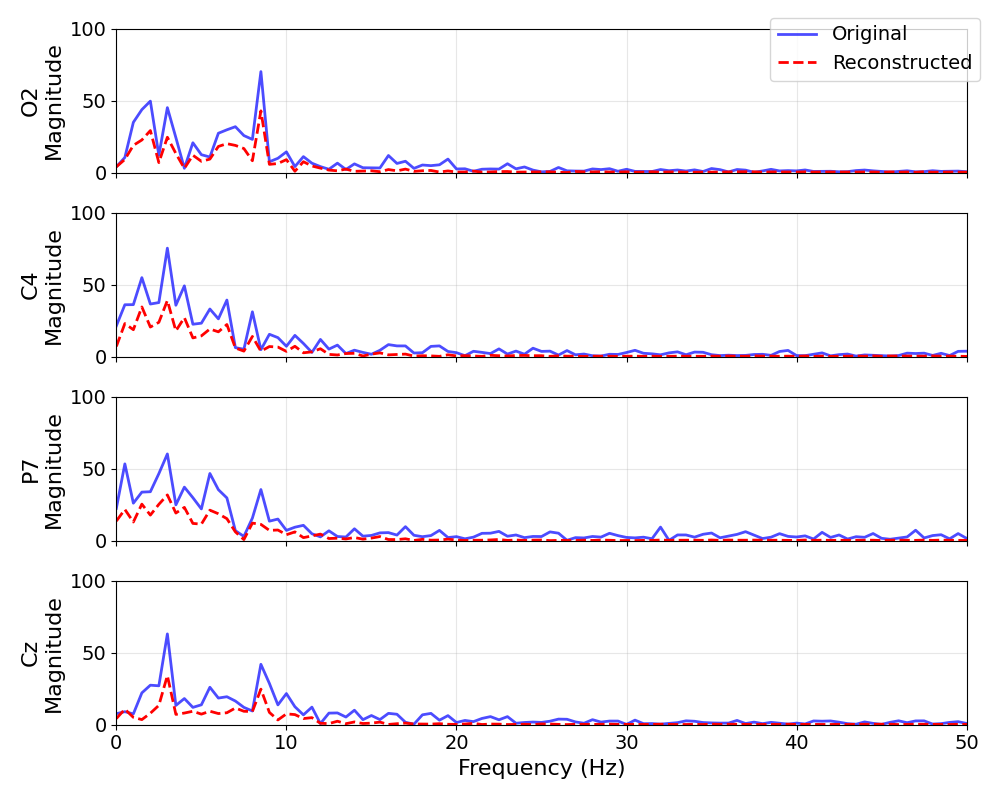}
			\caption{Example of the Fourier transform of the original and reconstructed time series using DCAE\textsubscript{ts-loss} for selected channels.}
			\label{fig:fourier}
		\end{figure}
		
		To evaluate the impact of different loss functions on autoencoder performance, three distinct DCAE models were trained using the loss functions specified in Sec. \ref{model_training}. The first model, DCAE\textsubscript{ts-loss}, was trained exclusively using the time series reconstruction loss. The second model, DCAE\textsubscript{ts-ft-loss}, incorporated both the time series and Fourier transform-based loss, as defined in Eq. \ref{loss_ts_ft}. The third model, DCAE\textsubscript{ts-stft-loss}, utilized a combination of time series and spectrogram-based loss, as formulated in Eq. \ref{loss_ts_spectro}. 
		
		Fig. \ref{fig:eeg} illustrates an original EEG time series signal along with its reconstruction by DCAE\textsubscript{ts-loss}. The blue line represents the original signal over a two-second interval for a subset of four different EEG channels, located at distinct scalp positions (\textit{O2}, \textit{C4}, \textit{P7} and \textit{Cz}). The red line is the reconstructed signal. Although the reconstruction is not perfect, the overall temporal characteristics of the EEG signal are well preserved by the autoencoder.
		
		Fig. \ref{fig:fourier} presents the Fourier transform of the original and reconstructed EEG signals from Fig. \ref{fig:eeg}, obtained using the same DCAE model. The Fourier transform is displayed without normalization. The EEG signal contains a predominance of low-frequency components, while higher frequencies above 20~Hz appear less frequently. Low-frequency components (0.5-8 Hz) are often associated with normal background brain activity, whereas epileptic seizures frequently manifest through abnormal activity in higher-frequency bands \cite{Niedermeyer2017}. As illustrated in Fig. \ref{fig:fourier}, the DCAE model demonstrated its capacity to reconstruct the majority of the frequency components well. The first row of the image depicting the Electrode "O2" reveals that the AE models the frequency peaks, e.g. at approximately 8 Hz, precisely.
		
		Table \ref{table_results} summarizes the quantitative evaluation of the trained DCAE models. 
		
		The first row shows the reconstruction error for the time series signal. As expected, the DCAE\textsubscript{ts-loss} achieves the lowest reconstruction error, indicating the highest fidelity in time domain reconstruction. The DCAE\textsubscript{ts-stft-loss} follows closely with only minor differences, while the DCAE\textsubscript{ts-ft-loss} demonstrates significantly lower reconstruction accuracy in the time domain. 
		
		The second row of the Tab. \ref{table_results} presents the reconstruction error in the frequency domain, computed as the difference between the Fourier-transformed original and reconstructed time series signals. In contrast to the time series error, the DCAE\textsubscript{ts-ft-loss} gives the lowest error and the DCAE\textsubscript{ts-loss} the highest.

		\begin{table}[t!]
			\caption{Validation of autoencoder reconstruction performance, including the time series reconstruction error, and frequency spectrum error.}
			\label{table_results}
			\centering
			\small 
			\setlength{\tabcolsep}{4pt} 
			\renewcommand{\arraystretch}{0.9} 
			\begin{tabular}{l c c c}
				& \makecell[l]{DCAE \\ ts-loss} & \makecell[l]{DCAE \\ ts-ft-loss} & \makecell[l]{DCAE \\ ts-stft-loss} \\
				\midrule
				\midrule
				MAE time (TS) & \textbf{0.107571} & 0.126109 & 0.109854 \\
				\midrule
				MAE frequency (FT) & 0.042120 &\textbf{0.038266} & 0.040176 \\
			\end{tabular}
		\end{table}
	
	\section{DISCUSSION}
		
		This study investigates whether a Deep Convolutional Autoencoder (DCAE) can learn, in an unsupervised manner, the characteristic features of a time series, including both spectral and temporal components. In particular, we focus on EEG data, which is inherently non-linear, non-stationary and dynamic. Our results show that pure amplitude reconstruction performs well when the Mean Absolute Error (MAE) of the reconstruction is used as the loss function, as shown in Fig.~\ref{fig:eeg}. However, the reconstruction struggles to preserve spectral features, as shown in Fig.~\ref{fig:fourier}. 
		
		To address these limitations, an additional autoencoder was trained with an augmented loss function incorporating the Fourier transform. This approach aims to better preserve the spectral properties of the signal. As shown in Tab.~\ref{table_results}, the Fourier reconstruction error improved compared to the DCAE\textsubscript{ts-loss} model. However, a notable trade-off was observed: the time series reconstruction error increased. This effect is due to the nature of the Fourier transform, which does not provide information about the exact temporal location of frequency components. As a result, the model successfully reconstructs higher frequency components, but they are not necessarily placed at the correct temporal locations.
		
		To mitigate this issue, a third model, DCAE\textsubscript{ts-stft-loss}, has been introduced that incorporates the Short-Time Fourier Transform (STFT) loss in addition to the standard reconstruction error. The advantage of the STFT over the Fourier transform is that it preserves both the spectral information and its temporal mapping, allowing the autoencoder to better detect when higher reconstruction errors occur. As shown in Tab.~\ref{table_results}, this model does not achieve the best performance in either pure amplitude reconstruction or spectral reconstruction, but it performs consistently well in both domains, achieving results close to the best models in each respective category.
		
		This study provides insights into how deep learning models encode temporal and spectral EEG features, highlighting the importance of explicitly incorporating frequency domain information. 
		
		Many existing EEG-based deep learning models rely on raw time series data, potentially neglecting spectral information. Researchers developing deep learning-based seizure detection models can benefit from these findings by considering hybrid loss functions that preserve both time and frequency characteristics. Alternatively, the spectral information can be included in addition to time series as input to any deep learning algorithm. The findings support the development of more robust seizure detection pipelines, particularly for models that rely on frequency domain features (e.g., spectral energy in specific EEG bands). 
		
		The work demonstrates that standard time series autoencoders struggle to preserve spectral properties, highlighting the need for tailored loss functions. Future studies on unsupervised EEG feature learning can use this approach to extract more informative latent representations.
	
	\section{CONCLUSION}
	
		This paper presents the development of a Deep Convolutional Autoencoder with different loss functions to investigate the impact of reconstruction objectives on EEG signal representation. By incorporating loss functions that go beyond plain time series reconstruction, the focus was shifted to preserving both the temporal and spectral properties of the EEG signal. Key findings indicate that while the general waveform characteristics of the EEG signal are well preserved, distinct frequency properties are not inherently kept. 
		This suggests that if frequency components are critical for given analysis task, e.g., EEG data, they must be explicitly considered during model design, either as part of the loss function or as an additional input representation in convolutional layers.
		
		The proposed method has implications for seizure detection and other EEG-based classification tasks. Traditional deep learning models do not automatically prioritize frequency domain information, despite its importance in EEG analysis. By demonstrating how different loss functions impact reconstruction, this work highlights the need for explicit frequency preservation in deep learning-based EEG analysis. 
		
		As a future direction, the proposed autoencoder can serve as a preprocessing step for seizure detection models, improving the feature extraction pipeline by ensuring relevant time series and frequency information in the latent space.

	\bibliographystyle{IEEEtran}
	\bibliography{IEEEabrv, embc_bib.bib}

\begin{thebibliography}{10}
\providecommand{\url}[1]{#1}
\csname url@samestyle\endcsname
\providecommand{\newblock}{\relax}
\providecommand{\bibinfo}[2]{#2}
\providecommand{\BIBentrySTDinterwordspacing}{\spaceskip=0pt\relax}
\providecommand{\BIBentryALTinterwordstretchfactor}{4}
\providecommand{\BIBentryALTinterwordspacing}{\spaceskip=\fontdimen2\font plus
\BIBentryALTinterwordstretchfactor\fontdimen3\font minus
  \fontdimen4\font\relax}
\providecommand{\BIBforeignlanguage}[2]{{%
\expandafter\ifx\csname l@#1\endcsname\relax
\typeout{** WARNING: IEEEtran.bst: No hyphenation pattern has been}%
\typeout{** loaded for the language `#1'. Using the pattern for}%
\typeout{** the default language instead.}%
\else
\language=\csname l@#1\endcsname
\fi
#2}}
\providecommand{\BIBdecl}{\relax}
\BIBdecl

\bibitem{WHO2024}
\BIBentryALTinterwordspacing
WHO. (2024, Feb.) \BIBforeignlanguage{en}{Epilepsy}. World Health Organization.
  [Online]. Available:
  \url{https://www.who.int/news-room/fact-sheets/detail/epilepsy}
\BIBentrySTDinterwordspacing

\bibitem{HerbozoContreras2024}
L.~F. Herbozo~Contreras, Z.~Huang, L.~Yu, A.~Nikpour, and O.~Kavehei,
  ``Biological plausible algorithm for seizure detection: Toward ai-enabled
  electroceuticals at the edge,'' \emph{APL Machine Learning}, vol.~2, no.~2,
  Apr. 2024.

\bibitem{Ma2024}
H.~Ma, Y.~Wu, Y.~Tang, R.~Chen, T.~Xu, and W.~Zhang, ``Parallel dual-branch
  fusion network for epileptic seizure prediction,'' \emph{Computers in Biology
  and Medicine}, vol. 176, p. 108565, Jun. 2024.

\bibitem{Ebner2011}
A.~Ebner and T.~Bast, \emph{EEG}, 2nd~ed., ser. Thieme e-book library.\hskip
  1em plus 0.5em minus 0.4em\relax Stuttgart [u.a.]: Thieme, 2011.

\bibitem{Mansilla2024}
D.~Mansilla, J.~Tveit, H.~Aurlien, T.~Avigdor, V.~Ros‐Castello, A.~Ho,
  C.~Abdallah, J.~Gotman, S.~Beniczky, and B.~Frauscher, ``Generalizability of
  electroencephalographic interpretation using artificial intelligence: An
  external validation study,'' \emph{Epilepsia}, vol.~65, no.~10, pp.
  3028--3037, Aug. 2024.

\bibitem{Dan2024}
J.~Dan, U.~Pale, A.~Amirshahi, W.~Cappelletti, T.~M. Ingolfsson, X.~Wang,
  A.~Cossettini, A.~Bernini, L.~Benini, S.~Beniczky, D.~Atienza, and P.~Ryvlin,
  ``Szcore: A seizure community open-source research evaluation framework for
  the validation of eeg-based automated seizure detection algorithms,'' 2024.

\bibitem{Sousa2024}
A.~M.~A. de~Sousa, M.~J. van Putten, S.~van~den Berg, and M.~Amir~Haeri,
  ``Detection of interictal epileptiform discharges with semi-supervised deep
  learning,'' \emph{Biomedical Signal Processing and Control}, vol.~88, p.
  105610, Feb. 2024.

\bibitem{Zhou2024}
Q.~Zhou, S.~Zhang, Q.~Du, and L.~Ke, ``Rihanet: A residual-based inception with
  hybrid-attention network for seizure detection using eeg signals,''
  \emph{Computers in Biology and Medicine}, vol. 171, p. 108086, Mar. 2024.

\bibitem{DeBrabandere2023}
A.~De~Brabandere, C.~Chatzichristos, W.~Van~Paesschen, M.~De~Vos, and J.~Davis,
  ``Detecting epileptic seizures using hand-crafted and automatically
  constructed eeg features,'' \emph{IEEE Transactions on Biomedical
  Engineering}, pp. 1--10, 2023.

\bibitem{Kerr2024}
W.~T. Kerr, K.~N. McFarlane, and G.~Figueiredo~Pucci, ``The present and future
  of seizure detection, prediction, and forecasting with machine learning,
  including the future impact on clinical trials,'' \emph{Frontiers in
  Neurology}, vol.~15, Jul. 2024.

\bibitem{GowthamReddy2024}
G.~R. N., S.~R. Hait, D.~Guha, and M.~Mahadevappa, ``Classification of
  epileptic eeg signals with the utilization of bonferroni mean based fuzzy
  pattern tree,'' \emph{Expert Systems with Applications}, vol. 239, p. 122424,
  Apr. 2024.

\bibitem{Li2024}
Y.~Li, Y.~Yang, S.~Song, H.~Wang, M.~Sun, X.~Liang, P.~Zhao, B.~Wang, N.~Wang,
  Q.~Sun, and Z.~Han, ``Multi-branch fusion graph neural network based on
  multi-head attention for childhood seizure detection,'' \emph{Frontiers in
  Physiology}, vol.~15, Oct. 2024.

\bibitem{Pedoeem2022}
J.~Pedoeem, G.~Bar~Yosef, S.~Abittan, and S.~Keene, \emph{TABS: Transformer
  Based Seizure Detection}.\hskip 1em plus 0.5em minus 0.4em\relax Springer
  International Publishing, 2022, pp. 133--160.

\bibitem{Hassan2024}
M.~M. Hassan, R.~Haque, S.~M.~S. Islam, H.~Meshref, R.~Alroobaea, M.~Masud, and
  A.~K. Bairagi, ``Neurowave-net: Enhancing epileptic seizure detection from
  eeg brain signals via advanced convolutional and long short-term memory
  networks,'' \emph{AIMS Bioengineering}, vol.~11, no.~1, pp. 85--109, 2024.

\bibitem{Amin2016}
S.~Amin and A.~M. Kamboh, ``A robust approach towards epileptic seizure
  detection,'' in \emph{2016 IEEE 26th International Workshop on Machine
  Learning for Signal Processing (MLSP)}.\hskip 1em plus 0.5em minus
  0.4em\relax IEEE, Sep. 2016.

\bibitem{Peh2023}
W.~Y. Peh, P.~Thangavel, Y.~Yao, J.~Thomas, Y.-L. Tan, and J.~Dauwels,
  ``Six-center assessment of cnn-transformer with belief matching loss for
  patient-independent seizure detection in eeg,'' \emph{International Journal
  of Neural Systems}, vol.~33, no.~03, Feb. 2023.

\bibitem{Tsai2023}
C.-W. Tsai, R.~Jiang, L.~Zhang, M.~Zhang, and J.~Yoo,
  ``Seizure-cluster-inception cnn (scicnn): A patient-independent epilepsy
  tracking soc with 0-shot-retraining,'' \emph{IEEE Transactions on Biomedical
  Circuits and Systems}, vol.~17, no.~6, pp. 1202--1213, Dec. 2023.

\bibitem{Gotman1976}
J.~Gotman and P.~Gloor, ``Automatic recognition and quantification of
  interictal epileptic activity in the human scalp eeg,''
  \emph{Electroencephalography and Clinical Neurophysiology}, vol.~41, no.~5,
  pp. 513--529, Nov. 1976.

\bibitem{Stiehl2023}
A.~Stiehl, M.~Flammer, F.~Anselstetter, N.~Ille, H.~Bornfleth,
  S.~Geißelsöder, and C.~Uhl, ``Topological analysis of low dimensional phase
  space trajectories of high dimensional eeg signals for classification of
  interictal epileptiform discharges,'' in \emph{2023 IEEE International
  Conference on Acoustics, Speech, and Signal Processing Workshops
  (ICASSPW)}.\hskip 1em plus 0.5em minus 0.4em\relax IEEE, Jun. 2023, pp. 1--5.

\bibitem{Tong2024}
P.~F. Tong, H.~X. Zhan, and S.~X. Chen, ``Ensembled seizure detection based on
  small training samples,'' \emph{IEEE Transactions on Signal Processing},
  vol.~72, pp. 1--14, 2024.

\bibitem{Gao2020}
Y.~Gao, B.~Gao, Q.~Chen, J.~Liu, and Y.~Zhang, ``Deep convolutional neural
  network-based epileptic electroencephalogram (eeg) signal classification,''
  \emph{Frontiers in Neurology}, vol.~11, May 2020.

\bibitem{SilvaLourenco2021}
C.~da~Silva~Lourenço, M.~C. Tjepkema-Cloostermans, and M.~J. van Putten,
  ``Machine learning for detection of interictal epileptiform discharges,''
  \emph{Clinical Neurophysiology}, vol. 132, no.~7, pp. 1433--1443, Jul. 2021.

\bibitem{Wen2018}
T.~Wen and Z.~Zhang, ``Deep convolution neural network and autoencoders-based
  unsupervised feature learning of eeg signals,'' \emph{IEEE Access}, vol.~6,
  pp. 25\,399--25\,410, 2018.

\bibitem{Takahashi2020}
H.~Takahashi, A.~Emami, T.~Shinozaki, N.~Kunii, T.~Matsuo, and K.~Kawai,
  ``Convolutional neural network with autoencoder-assisted multiclass labelling
  for seizure detection based on scalp electroencephalography,''
  \emph{Computers in Biology and Medicine}, vol. 125, p. 104016, Oct. 2020.

\bibitem{Fu2024a}
Z.~Fu, H.~Zhu, Y.~Zhao, R.~Huan, Y.~Zhang, S.~Chen, and Y.~Pan, ``Gmaeeg: A
  self-supervised graph masked autoencoder for eeg representation learning,''
  \emph{IEEE Journal of Biomedical and Health Informatics}, vol.~28, no.~11,
  pp. 6486--6497, Nov. 2024.

\bibitem{Shah2018}
V.~Shah, E.~von Weltin, S.~Lopez, J.~R. McHugh, L.~Veloso, M.~Golmohammadi,
  I.~Obeid, and J.~Picone, ``The temple university hospital seizure detection
  corpus,'' \emph{Frontiers in Neuroinformatics}, vol.~12, Nov. 2018.

\bibitem{Golmohammadi2021}
M.~Golmohammadi, ``\BIBforeignlanguage{en}{Deep architectures for
  spatio-temporal sequence recognition with applications in automatic seizure
  detection},'' Ph.D. dissertation, Temple University, Department Electrical
  and Computer Engineering, 2021.

\bibitem{Goodfellow2016}
I.~Goodfellow, Y.~Bengio, and A.~Courville, \emph{Deep Learning}.\hskip 1em
  plus 0.5em minus 0.4em\relax MIT Press, 2016,
  \url{http://www.deeplearningbook.org}.

\bibitem{RobustScaler}
\BIBentryALTinterwordspacing
``\BIBforeignlanguage{en}{{RobustScaler}}.'' [Online]. Available:
  \url{https://scikit-learn/stable/modules/generated/sklearn.preprocessing.RobustScaler.html}
\BIBentrySTDinterwordspacing

\bibitem{Kingma2014}
D.~P. Kingma and J.~Ba, ``Adam: A method for stochastic optimization,'' 2014.

\bibitem{Niedermeyer2017}
E.~Niedermeyer, \emph{Niedermeyer’s Electroencephalography}, D.~L. Schomer
  and F.~H. Lopes~da Silva, Eds.\hskip 1em plus 0.5em minus 0.4em\relax Oxford
  University Press, Nov. 2017.

\bibitem{UrbinaFredes2024}
S.~Urbina~Fredes, A.~Dehghan~Firoozabadi, P.~Adasme, D.~Zabala-Blanco,
  P.~Palacios~Játiva, and C.~Azurdia-Meza, ``Enhanced epileptic seizure
  detection through wavelet-based analysis of eeg signal processing,''
  \emph{Applied Sciences}, vol.~14, no.~13, p. 5783, Jul. 2024.

\end{thebibliography}

\end{document}